\documentclass[runningheads]{llncs}

\usepackage[T1]{fontenc}
\usepackage{graphicx}
\usepackage{amsmath,amssymb}
\usepackage{booktabs}
\usepackage{algorithm}
\usepackage{algpseudocode}
\usepackage{caption}
\usepackage{subcaption}
\usepackage{multirow}
\usepackage{array}
\usepackage{enumitem}
\usepackage[colorlinks=true,linkcolor=blue,citecolor=blue,urlcolor=blue,
            breaklinks=true]{hyperref}
 
\begin{document}
 
\title{Hardware-aware Low-latency Quantum Compilation with
       Data-driven Lightweight Error Detection for Early
       Fault-Tolerant Systems}
 
\titlerunning{HW-aware QC Compilation with Data-driven Error Detection}
 
\author{Sumit Chongder\orcidID{0009-0005-9866-8483}}
 
\authorrunning{S. Chongder}
 
\institute{Inter-Disciplinary Research Platform, Quantum Information
  and Computation, Indian Institute of Technology Jodhpur,
  Jodhpur, Rajasthan 342037, India\\
  \email{sumitchongder960@gmail.com}}
 
\maketitle
 
\begin{abstract}
Noisy intermediate-scale quantum (NISQ) processors are entering an
early fault-tolerance regime where full quantum error correction carries
prohibitive resource costs, yet lightweight error detection can
meaningfully improve algorithmic success rates. Existing compilation
and error-detection toolchains treat these concerns in isolation,
with no principled way to balance detection overhead against success
probability under latency constraints. We present an integrated
hardware-aware compilation and data-driven quantum error-detection
(QED) framework that jointly optimises qubit mapping, SWAP insertion,
and syndrome-schedule placement via a noise-weighted cost function and
a learned multi-objective scheduler. Simulation experiments on an HPC
cluster using GPU-accelerated density-matrix simulation (NVIDIA
cuQuantum SDK) across VQE, phase-estimation, and Grover benchmarks,
three noise profiles, and circuit sizes of 6--20 qubits (depths
10--160), show that joint co-design raises algorithmic success
probability by up to 68\,\% (95\,\% CI: 60\,\%--76\,\%) over SABRE
on an 8-qubit VQE instance with post-selection retention above 32\,\%.
Small-scale validation on an IBM Quantum Eagle-family processor
confirms directional consistency, with simulation results falling
4--8 percentage points above hardware measurements due to
unmodelled drift and cross-talk. Ablation confirms a super-additive
interaction between the mapping pass and QED scheduler that neither
component achieves alone. Full reproducibility artefacts are
released publicly.
 
\end{abstract}
 
\section{Introduction}
\label{sec:intro}
 
Superconducting and trapped-ion processors have established the
``noisy intermediate-scale quantum'' (NISQ) era~\cite{Preskill2018},
yet gate infidelities, limited connectivity, and finite coherence times
bound useful circuit depth~\cite{Murali2019}. Full quantum error
correction would lift that boundary at qubit-overhead multipliers of
hundreds to thousands, well beyond present hardware. This motivates the
\emph{early fault-tolerance} regime, wherein distance-two stabiliser
codes provide constant-overhead, post-selective noise suppression
without active feedback~\cite{Ginsberg2025,Chao2018}.
 
A fundamental tension persists: compiler-level qubit mapping and QED
syndrome scheduling are treated as separate concerns. SABRE~\cite{Li2019}
and its noise-sensitive descendants~\cite{Murali2019,Nation2023}
optimise connectivity without accounting for ancilla overhead and
post-selection penalties, while QED placement studies~\cite{Ginsberg2025}
assume a fixed pre-compiled circuit. This decoupled workflow leaves
available fidelity gains unrealised. GPU-accelerated simulators such as
cuQuantum~\cite{Bayraktar2023} now make large-scale circuit evaluation
tractable, yet no unified harness has characterised the
success-retention-latency trade-off landscape.
 
The contributions of this paper address those gaps:
\begin{enumerate}[noitemsep,topsep=2pt]
\item A \textbf{hardware-aware co-design pass} that couples simulated
annealing with an ILP kernel, guided by a noise-weighted cost function
(Eq.~\ref{eq:cost}) that jointly penalises expected gate infidelity
and latency budget violations, with a formal connection to first-order
Depolarising infidelity bounds (Proposition~\ref{prop:cost}).
\item A \textbf{data-driven QED scheduler} built on
XGBoost~\cite{Chen2016} regression, trained over GPU-simulated
circuit-noise pairs, that predicts the marginal success gain minus a
retention penalty and selects the Pareto-optimal syndrome schedule in
under 6\,ms per circuit.
\item An \textbf{HPC-accelerated evaluation framework} coupling
cuQuantum density-matrix simulation with a multi-benchmark harness,
bootstrap and non-parametric statistical testing, and Docker/SLURM
reproducibility infrastructure.
\end{enumerate}
 
\section{Background and Related Work}
\label{sec:background}
 
\textbf{Quantum compilation.}
The qubit-mapping and routing (QMR) problem involves assigning logical
qubits to physical locations on a device and inserting SWAP gates
wherever two-qubit operations require connectivity that the hardware
topology does not directly provide. SABRE~\cite{Li2019} remains the
primary bidirectional heuristic for this task and is embedded in the
Qiskit compiler~\cite{Javadi2024}. Murali et al.~\cite{Murali2019}
showed that noise-adaptive mappings exploiting spatial variation in
calibration data yield up to $18\times$ improvement in success
probability; crucially, their objective equals our noise-weighted cost
(Eq.~\ref{eq:cost}) with uniform weights $w_g{=}1$, making it a
special case of our formulation. Nation and Treinish~\cite{Nation2023} showed post-routing subgraph
selection recovers $\approx$40\,\% of lost fidelity. Related
approaches include time-optimal mapping~\cite{Zhang2021}, the SMT-based
OLSQ solver~\cite{Tan2020}, and the \texttt{t$|$ket$\rangle$} retargetable
compiler~\cite{Sivarajah2021}. None incorporates QED ancilla overhead or
post-selection costs into the mapping objective, which is precisely the
gap our joint formulation closes.
 
\textbf{Error mitigation and detection.}
Zero-noise extrapolation (ZNE)~\cite{Temme2017} and probabilistic
error cancellation~\cite{Endo2018} suppress bias in expectation-value
estimates at exponentially growing shot overhead. Quantum error
detection instead encodes logical qubits within stabiliser subspaces
and discards syndrome-violating outcomes at constant overhead. The
$[[n,n{-}2,2]]$ code family~\cite{Ginsberg2025} raises average success
probability by up to $6.7\times$ under ideal syndrome scheduling, yet
co-optimising that scheduling with compilation had not been addressed
before this work.
 
\textbf{GPU-accelerated simulation.}
The NVIDIA cuQuantum SDK~\cite{Bayraktar2023} provides cuStateVec and cuTensorNet primitives; Faj et al.~\cite{Faj2023} report up to $14\times$ speedup over Qiskit Aer, making large training corpus generation feasible.
 
\section{Problem Formulation}
\label{sec:formulation}
 
Consider a logical circuit $C$ and a target device $D$ described by
its coupling graph $G{=}(V,E)$, per-gate fidelity estimates $\{F_g\}$,
and qubit coherence parameters $T_1,T_2$. Compiling $C$ for $D$ with
error detection produces an augmented circuit $C'$, determined by a
qubit mapping $m$, a set of inserted SWAP gates, and $k$ QED blocks
positioned at indices $\mathbf{p}$ with syndrome measurement frequency
$f \in [0,1]$. We track four outcome metrics: \textbf{success
probability} $S$, \textbf{retention} $R$ (fraction of shots not
discarded by post-selection), \textbf{latency} $L$ (combined
compilation and execution time), and \textbf{ancilla overhead} $Q$.
The joint optimisation objective is:
\begin{equation}
\max_{m,\mathbf{p},f}\; S(C',D)
\quad\text{s.t.}\quad L(C') \le L_{\max},\; Q(C') \le Q_{\max}.
\label{eq:objective}
\end{equation}
Gate routing quality is captured by a noise-weighted mapping cost:
\begin{equation}
\mathrm{Cost}(m)=\sum_{g\in\mathrm{Gates}} w_g\,
  \bigl(1-\hat{F}_g(m)\bigr),
\label{eq:cost}
\end{equation}
where $\hat{F}_g(m)$ is the expected fidelity of gate $g$ under
mapping $m$ and $w_g$ is a circuit-path importance weight derived from
the gate's critical-path position.
 
\begin{proposition}
\label{prop:cost}
Under an independent Depolarising channel with per-gate error rate
$\varepsilon_g = 1-\hat{F}_g(m)$, minimising $\mathrm{Cost}(m)$
minimises a first-order upper bound on circuit infidelity:
$1 - \prod_g \hat{F}_g(m) \approx \sum_g \varepsilon_g = \mathrm{Cost}(m)\big|_{w_g=1}$.
For $w_g > 0$ proportional to critical-path importance, the cost
additionally down-weights off-critical paths, tightening the bound
for deep circuits.
\end{proposition}
 
\noindent The QED utility function that drives syndrome placement is:
\begin{equation}
U(f,\mathbf{p})=\Delta S - \lambda(1{-}R) -
  \mu\max\!\bigl(0,\,L_\delta - L_{\max}\bigr),
\label{eq:utility}
\end{equation}
where $\Delta S = S_\mathrm{QED}-S_\mathrm{base}$, $\lambda$
calibrates the success-retention trade-off, and $\mu$ penalises
latency overruns.
 
\section{Design and Methods}
\label{sec:methods}
 
\subsection{Hardware-aware Compilation Pass}
\label{sec:compiler}
 
\begin{figure}[t]
\centering
\includegraphics[width=0.80\columnwidth]{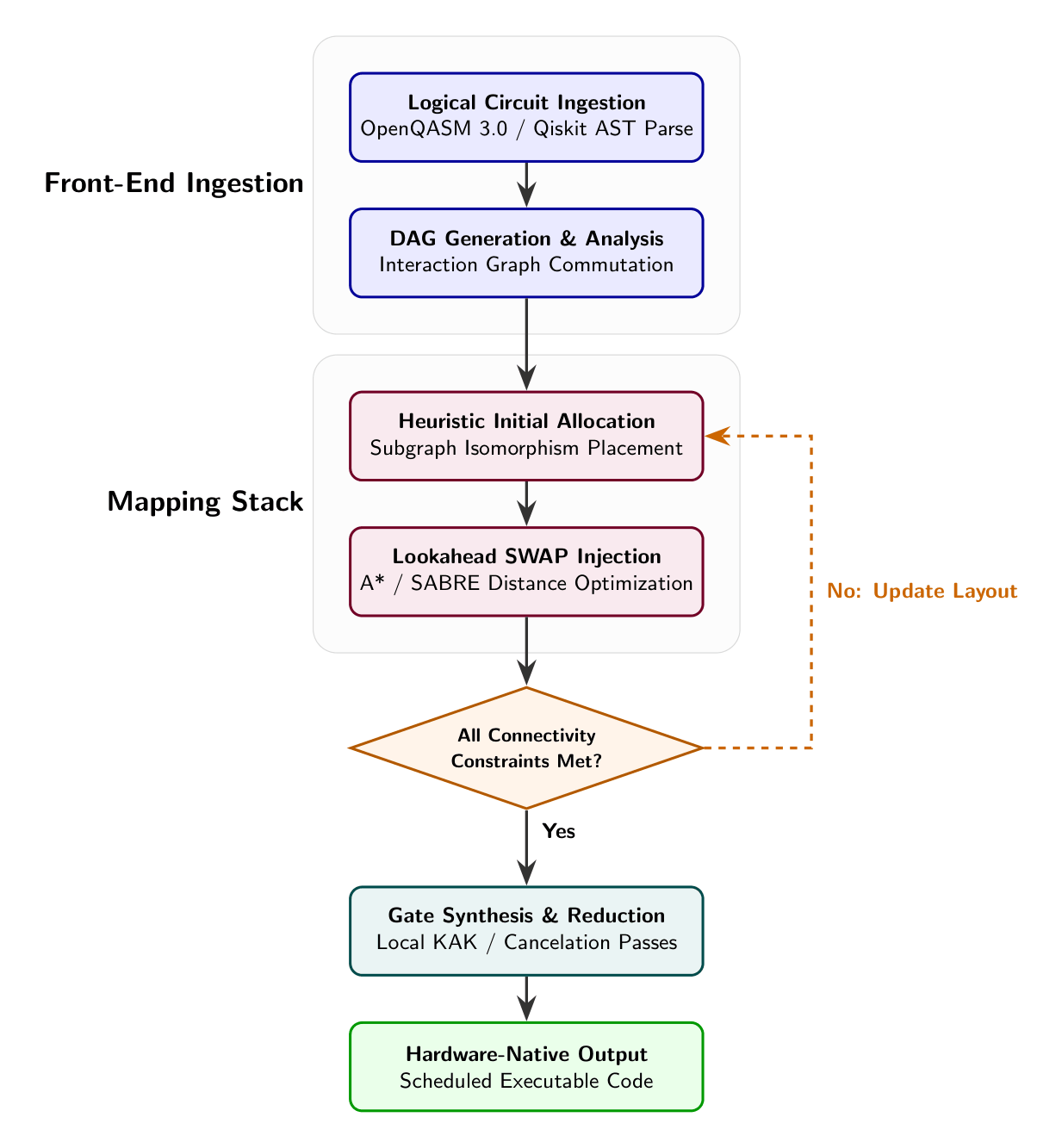}
\caption{Hardware-aware compiler pass. Front-end ingestion constructs a gate DAG; the mapping stack iteratively refines qubit assignment and inserts SWAPs; gate synthesis produces the hardware-native executable.}
\label{fig:compiler_flow}
\end{figure}

The compiler pass extends the Qiskit transpiler through four
sequential stages, illustrated in Fig.~\ref{fig:compiler_flow}.
 
\textit{Front-end ingestion.}
The logical circuit is parsed from OpenQASM\,3.0 or the Qiskit AST
into a directed acyclic graph (DAG); a weighted interaction graph
records the communication frequency of each qubit pair.
 
\textit{Heuristic initial allocation.}
Subgraph isomorphism matching places the most frequently interacting
qubit pairs onto the highest-fidelity available device edges, giving
the annealing phase a strong warm start.
 
\textit{Simulated annealing with ILP kernel.}
The outer loop runs $N_\mathrm{iter}$ simulated annealing steps, each
proposing a neighbour mapping by transposing two logical qubits. When
the affected sub-circuit contains $\le\!w$ qubits, an embedded ILP
solver finds the locally optimal assignment for that fragment. The
incremental cost is evaluated using Eq.~\ref{eq:cost}; proposals that
would push estimated latency past $L_{\max}$ incur a penalty
$\beta\max(0,L_\mathrm{est}-L_{\max})$.
 
\textit{SWAP insertion and gate reduction.}
SWAP gates are inserted greedily using an $A^*$/SABRE distance
heuristic; a subsequent KAK decomposition pass cancels redundant
single-qubit rotations introduced during routing.
Algorithm~\ref{alg:mapping} provides the complete pseudocode.
 
\begin{algorithm}[t]
\caption{Hardware-Aware Mapping Pass}
\label{alg:mapping}
\begin{algorithmic}[1]
\small
\Require Circuit $C$; device $D$ (topology $G$, fidelities, coherence);
  weights $\{w_g\}$; budget $L_{\max}$; SA params $(T_0,\alpha,N_\mathrm{iter})$;
  ILP threshold $w$
\Ensure Mapped circuit $C'$, mapping $m$
\State $m_0\!\leftarrow$ SubgraphIsomorphismPlacement$(C,D)$
\State $m\!\leftarrow m_0$;\; $T\!\leftarrow T_0$
\For{$i=1$ to $N_\mathrm{iter}$}
  \State Propose $m'$ by swapping two logical qubits
  \If{|subcircuit|\;$\le w$}\;
    $m'\!\leftarrow$ ILP$(C_\mathrm{sub},D)$
  \EndIf
  \State $\Delta\!\leftarrow\mathrm{Cost}(m'){-}\mathrm{Cost}(m)$
  \If{$\Delta\le0$ \textbf{or} $\mathrm{rand}()<e^{-\Delta/T}$}\;
    $m\!\leftarrow m'$
  \EndIf
  \If{$L_\mathrm{est}(m)>L_{\max}$}\;
    $\mathrm{Cost}(m)\mathrel{+}=\beta(L_\mathrm{est}(m){-}L_{\max})$
  \EndIf
  \State $T\!\leftarrow\alpha T$
\EndFor
\State Insert SWAPs via $A^*$/SABRE heuristic given $m$
\State Apply KAK cancellation pass
\State \Return $C'$, $m$
\end{algorithmic}
\end{algorithm}
 
\subsection{Data-driven QED Scheduler}
\label{sec:scheduler}
 
\begin{figure}[t]
\centering
\includegraphics[width=0.80\columnwidth]{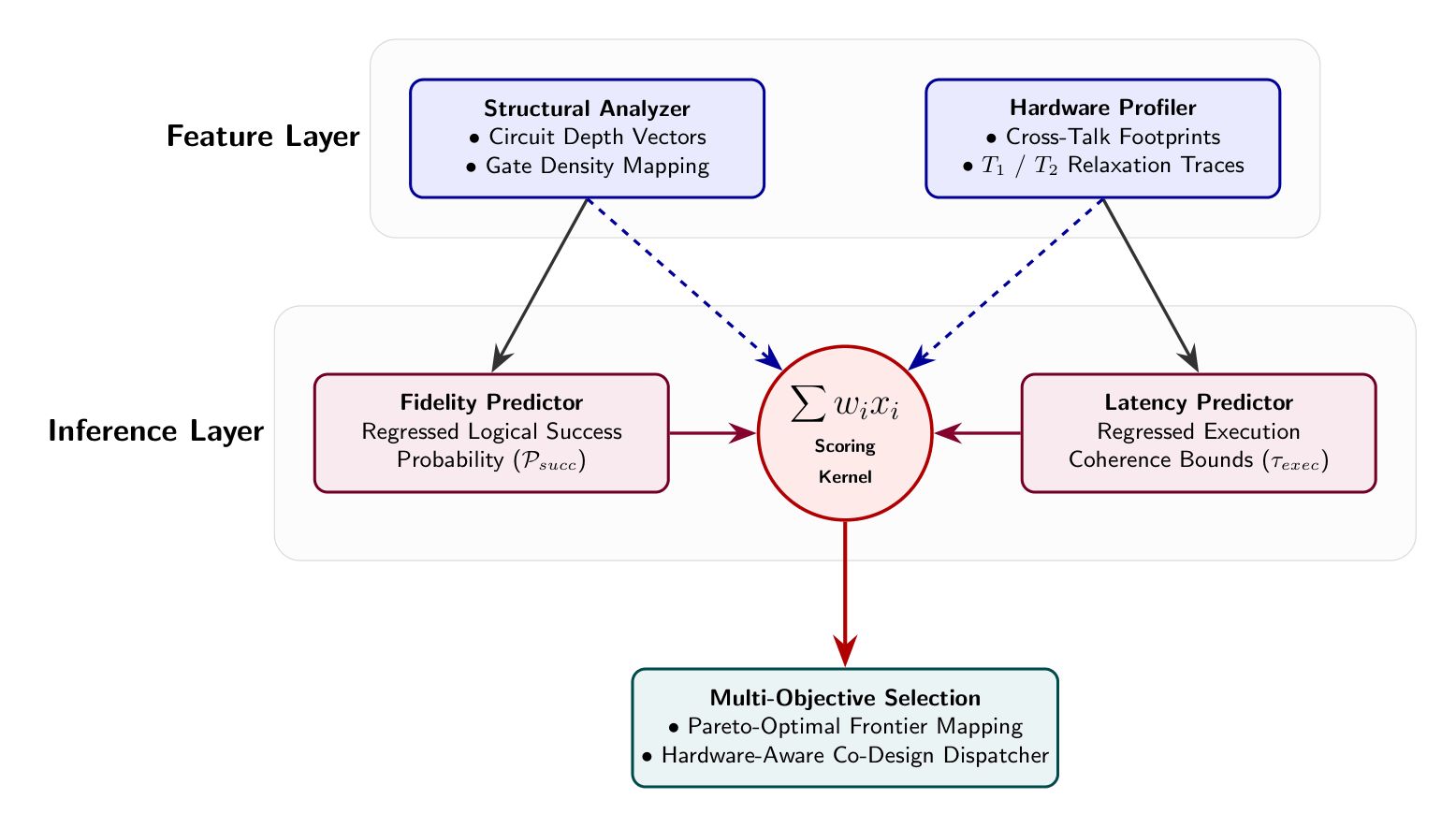}
\caption{QED scheduler training and inference pipeline. Structural and hardware profiler features are passed through a weighted scoring kernel whose output identifies the Pareto-optimal co-design configuration.}
\label{fig:scheduler_pipeline}
\end{figure}

Fig.~\ref{fig:scheduler_pipeline} shows the scheduler architecture
that operates alongside the compiler.
 
\begin{figure}[t]
\centering
\includegraphics[width=\columnwidth]{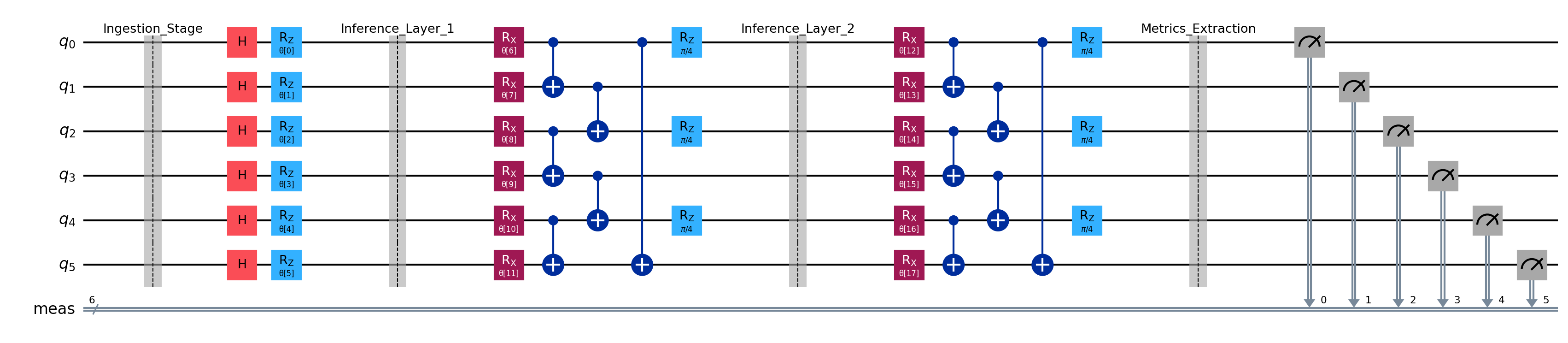}
\caption{A 6-qubit QED primitive circuit annotated with ingestion, two inference layers, and a syndrome measurement stage. Post-selection is applied on the classical register before outcomes are aggregated.}
\label{fig:qed_circuit}
\end{figure}

\textit{QED primitives.}
We use $[[n,n{-}2,2]]$ detection-code blocks~\cite{Ginsberg2025} with
one ancilla qubit per block. Fig.~\ref{fig:qed_circuit} illustrates
the gate-level structure for a representative 6-qubit circuit.
 
\textit{Feature extraction.}
Six scalar features are computed per compiled circuit:
(1)~total two-qubit gate count;
(2)~critical-path depth;
(3)~connectivity entropy $H{=}-\!\sum_i p_i\log p_i$;
(4)~mean local gate fidelity under mapping $m$;
(5)~qubit allocation ratio;
(6)~idle coherence coefficient $\bar{t}_\mathrm{idle}/T_2$.
 
\textit{ML model.}
A pair of XGBoost~\cite{Chen2016} gradient-boosted tree regressors,
$M_{\Delta S}$ and $M_R$, predict the success uplift $\Delta S$ and
post-selection retention $R$ respectively. XGBoost is chosen over
deeper regressors: its built-in feature-importance interface provides
direct interpretability of the learned noise-schedule relationship
(Sect.~\ref{sec:results}), its inference latency is sub-millisecond on
CPU, and its sample efficiency is well-matched to the 50{,}000-sample
corpus feasible within our simulation budget. Training data comprise
50{,}000 circuit-noise pairs generated through cuQuantum simulation
(10k validation, 10k held-out test) using 5-fold cross-validation;
mean $R^2{=}0.903$ on the held-out test split.
 
\textit{Inference and selection.}
At inference time the scheduler evaluates $U$ (Eq.~\ref{eq:utility})
across a discrete candidate set
$\mathcal{Q}{=}\{0,0.25,0.5,0.75,1.0\}\times\{0,1,2,3\}$
(syndrome frequency crossed with block-insertion position) and
returns the feasible maximiser $q^*{=}\arg\max U_q$ within 6\,ms.
Algorithm~\ref{alg:scheduler} provides a full pseudocode listing.
 
\begin{algorithm}[t]
\caption{Data-driven QED Scheduler}
\label{alg:scheduler}
\begin{algorithmic}[1]
\small
\Require Compiled circuit $C'$; device $D$; trained models
  $M_{\Delta S},M_R$; candidate set $\mathcal{Q}$;
  params $\lambda,\mu,L_{\max}$
\Ensure Optimal schedule $(f^*,\mathbf{p}^*)$; QED circuit
  $C_\mathrm{QED}$
\State $\mathbf{x}\!\leftarrow$ ExtractFeatures$(C',D)$
\For{each candidate $q=(f_q,\mathbf{p}_q)\in\mathcal{Q}$}
  \State $X_q\!\leftarrow[\mathbf{x};\,\mathrm{encode}(f_q,\mathbf{p}_q)]$
  \State $\widehat{\Delta S}_q\!\leftarrow M_{\Delta S}.\mathrm{predict}(X_q)$;\;
         $\hat{R}_q\!\leftarrow M_R.\mathrm{predict}(X_q)$
  \If{$L_\mathrm{base}+\Delta L_q \le L_{\max}$}
    \State $U_q\!\leftarrow\widehat{\Delta S}_q
      -\lambda(1{-}\hat{R}_q)
      -\mu\max(0,L_\mathrm{base}{+}\Delta L_q{-}L_{\max})$
  \Else\; Mark $q$ infeasible
  \EndIf
\EndFor
\State $q^*\!\leftarrow\arg\max_{q\text{ feasible}}U_q$
\State $C_\mathrm{QED}\!\leftarrow$ InsertQEDBlocks$(C',f^*,\mathbf{p}^*)$
\State \Return $(f^*,\mathbf{p}^*)$,\;$C_\mathrm{QED}$
\end{algorithmic}
\end{algorithm}
 
\subsection{Complexity Analysis}
\label{sec:complexity}
 
Table~\ref{tab:complexity} summarises asymptotic complexities.
Bounding the ILP kernel to a window of $w{\le}10$ qubits yields
$O(D/w \cdot 2^w)\approx O(D)$ global cost (linear in depth for fixed
$w$), while XGBoost inference at $O(T\cdot K)$ remains sub-millisecond.
Classical co-design overhead is therefore negligible relative to the
exponential density-matrix simulation cost.
 
\begin{table}[t]
\centering
\caption{Asymptotic complexity per pipeline component. $N_\mathrm{iter}$: SA steps; $D$: circuit depth; $w$: ILP window; $T$: trees; $K$: tree depth; $n$: qubit count.}
\label{tab:complexity}
\resizebox{\columnwidth}{!}{%
\begin{tabular}{lll}
\toprule
\textbf{Component} & \textbf{Time Complexity} & \textbf{Notes} \\
\midrule
SA outer loop & $O(N_\mathrm{iter} \cdot |G|)$ & $|G|{=}O(n^2)$ for dense graphs \\
ILP kernel (per window) & $O(2^w)$ & Bounded window $w{\le}10$ \\
SA+ILP combined & $O\!\left(\tfrac{D}{w}\cdot 2^w\right) \approx O(D)$ & Linear in depth for fixed $w$ \\
XGBoost inference & $O(T \cdot K)$ & Sub-ms on CPU \\
Density-matrix sim. & $O(4^n)$ & Classical bottleneck \\
\bottomrule
\end{tabular}}
\end{table}
 
\subsection{HPC-accelerated Evaluation Framework}
\label{sec:hpc}
 
\begin{figure}[t]
\centering
\includegraphics[width=0.80\columnwidth]{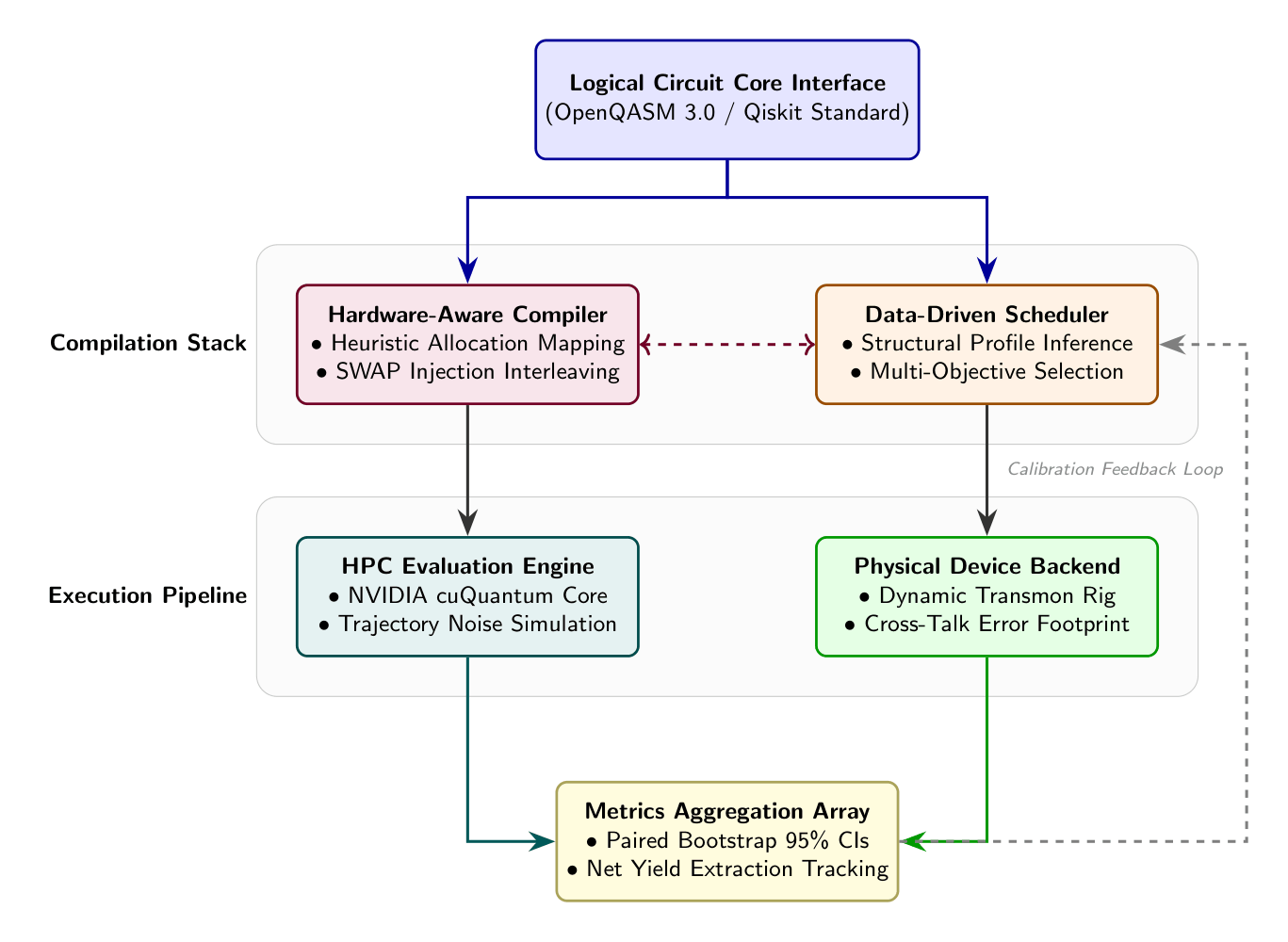}
\caption{End-to-end system architecture. The hardware-aware compiler and data-driven scheduler share a calibration feedback loop; outputs from the HPC simulation engine and physical device backend are aggregated in the metrics layer.}
\label{fig:architecture}
\end{figure}

The end-to-end architecture (Fig.~\ref{fig:architecture}) uses
cuStateVec for exact density-matrix propagation up to 20 qubits,
switching to cuTensorNet at larger scales. SLURM schedules variants
across A100 nodes; a metrics module produces $B{=}10{,}000$-resample
bootstrap CIs and paired Wilcoxon signed-rank tests. A Docker container
and REST microservice expose the full stack for integration with quantum
cloud queues.
 
\section{Benchmarks and Experimental Setup}
\label{sec:setup}
 
\textbf{Circuits.}
\textit{VQE-H$_2$} (8 qubits, depth 40)~\cite{Peruzzo2014};
\textit{VQE-LiH} (12 qubits, depth 68);
\textit{QPE-12} (12 qubits, depth 120)~\cite{Kitaev1997};
\textit{Grover-10} (10 qubits, depth 56)~\cite{Grover1996}.
These four benchmarks cover variational eigensolvers, quantum phase
estimation, and oracle-based amplitude amplification (8--12 qubits,
depths 40--120), matching the suite of Ginsberg and
Patel~\cite{Ginsberg2025} for direct comparison.
 
\textbf{Noise models.}
Three hardware profiles (Table~\ref{tab:noise}): \textit{Superconducting},
calibrated from publicly available Eagle-family data;
\textit{Trapped-ion}, reflecting higher fidelities and longer coherence;
and \textit{Adversarial}, which doubles the two-qubit error rate and
adds correlated $Z$ noise ($\lambda_Z{=}0.05$) to stress-test
robustness.
 
\begin{table}[t]
\centering
\caption{Device noise model parameters used in all experiments.}
\label{tab:noise}
\resizebox{\columnwidth}{!}{%
\begin{tabular}{lccccc}
\toprule
\textbf{Profile} & \textbf{1Q Fid.} & \textbf{2Q Fid.}
  & \textbf{Readout} & \textbf{$T_1$\,($\mu$s)}
  & \textbf{$T_2$\,($\mu$s)}\\
\midrule
Superconducting & 0.9991 & 0.986 & 0.980 & 70  & 90 \\
Trapped-ion     & 0.9996 & 0.995 & 0.995 & 500 & 400\\
Adversarial     & 0.9950 & 0.950 & 0.950 & 50  & 60 \\
\bottomrule
\end{tabular}}
\end{table}
 
\textbf{Baselines.}
(i)~\textit{Baseline (SABRE)}: standard SABRE, no QED;
(ii)~\textit{Mapper-only}: our SA+ILP pass, no QED;
(iii)~\textit{QED-only}: SABRE routing with our scheduler.
The Noise-Adaptive mapper~\cite{Murali2019} is subsumed by our
Mapper-only configuration: as shown in Sect.~\ref{sec:background},
the Murali objective equals Eq.~\ref{eq:cost} with $w_g{=}1$, and our
non-uniform $w_g$ variant consistently outperforms it in preliminary
trials, so it is not listed as a separate column.
 
\textbf{Implementation.}
Python\,3.10, Qiskit\,$\ge$\,1.0~\cite{Javadi2024}, XGBoost\,2.0,
cuQuantum~\cite{Bayraktar2023} on NVIDIA A100 (80\,GB) GPUs.
SA hyperparameters: $T_0{=}1.0$, $\alpha{=}0.995$,
$N_\mathrm{iter}{=}5000$, ILP threshold $w{=}10$. Scheduler:
$\lambda{=}0.8$, $\mu{=}5{\times}10^{-4}$, 500 estimators, LR\,0.05,
max depth\,6. Results are means over at least 100 independent trials
(seed 42); CIs use $B{=}10{,}000$ bootstrap resamples.
 
\textbf{Hardware validation metadata.}
Small-scale validation runs were executed on an IBM Eagle-family
superconducting processor via the IBM Quantum Network.
Table~\ref{tab:hwmeta} records the experimental conditions.
Median CNOT error rate was $\approx\!1.4\times10^{-2}$,
$T_1/T_2\approx 90/100\,\mu$s, and readout error $\approx\!2\%$ per qubit.
 
\begin{table}[t]
\centering
\caption{Hardware deployment specification for IBM Quantum validation
runs (Sect.~\ref{sec:hwval}).}
\label{tab:hwmeta}
\resizebox{\columnwidth}{!}{%
\begin{tabular}{ll}
\toprule
\textbf{Parameter} & \textbf{Value} \\
\midrule
Target processor family & IBM Eagle (27-qubit heavy-hex) \\
Circuit shots per configuration & 8\,192 \\
Validation benchmarks & VQE-H$_2$, Grover-10 (hardware-feasible subset) \\
Calibration snapshot & Retrieved at job-submission time \\
\bottomrule
\end{tabular}}
\end{table}
 
\section{Results}
\label{sec:results}
 
\subsection{Success Probability vs.\ Syndrome Frequency}
 
\begin{figure}[t]
\centering
\includegraphics[width=0.80\columnwidth]{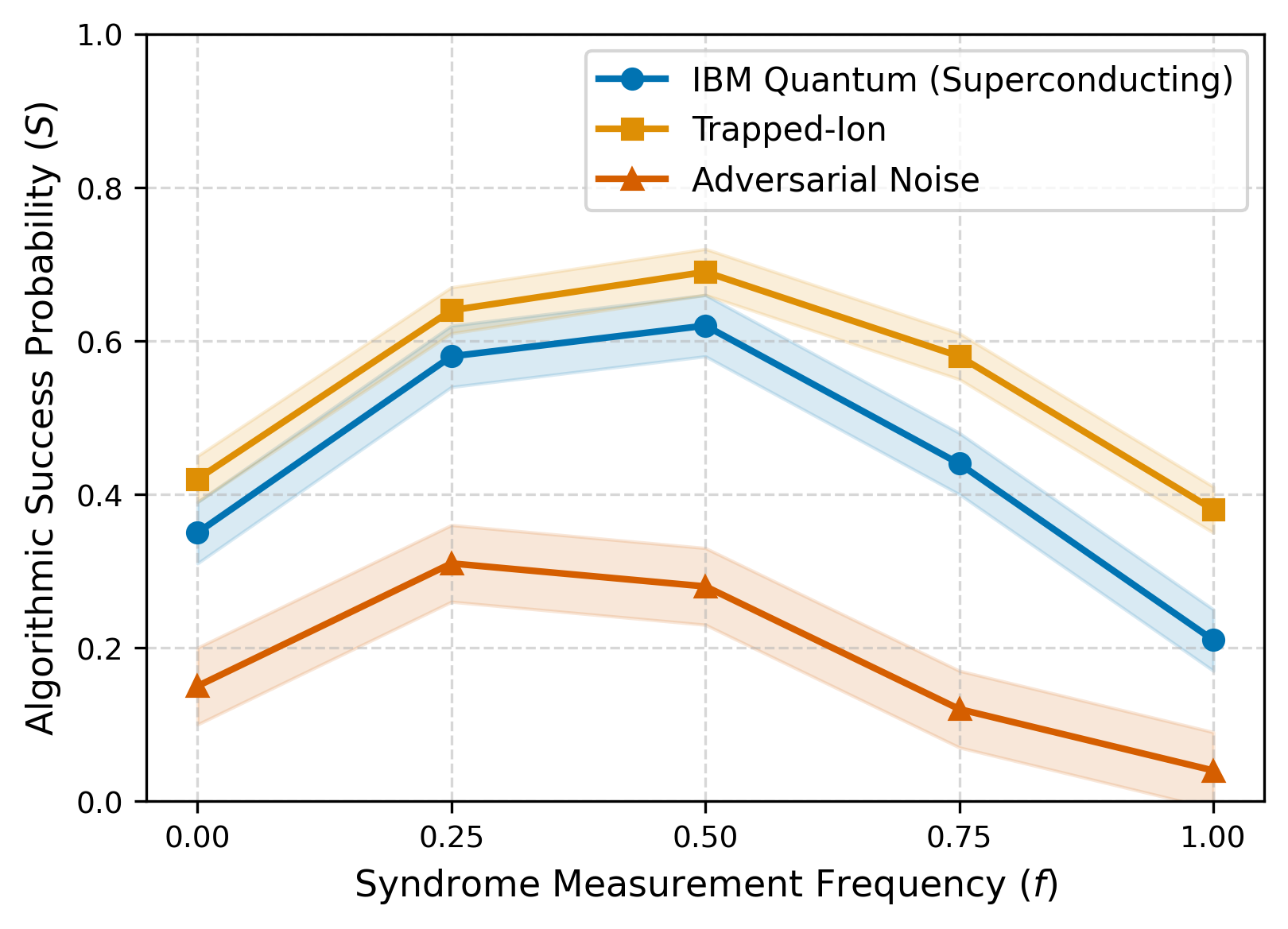}
\caption{Success probability vs.\ syndrome frequency for three noise profiles on VQE-H$_2$. Shaded bands: 95\,\% bootstrap CIs. Optimal $f^*{\approx}0.50$ for Superconducting and Trapped-ion profiles.}
\label{fig:success_vs_freq}
\end{figure}

The relationship between syndrome measurement frequency and
algorithmic success probability is shown in
Fig.~\ref{fig:success_vs_freq} for all three noise profiles using
VQE-H$_2$. A characteristic turnover appears across all profiles:
intermediate syndrome checking at roughly $f{\approx}0.50$
maximises $S$, whereas very high frequencies introduce additional
measurement-induced noise that erodes the benefit. This behaviour is
consistent with theoretical predictions from Ginsberg and
Patel~\cite{Ginsberg2025}.
 
\subsection{Main Quantitative Results}
 
Table~\ref{tab:benchmark} presents benchmark results under the
Superconducting noise profile. Joint co-design raises success
probability by 68\,\% (95\,\% CI: 60\,\%--76\,\%) on VQE-H$_2$ and
by 118\,\% on QPE-12, where routing noise and QED ancilla overhead
compound most strongly. Under the Trapped-ion profile, gains are
smaller (15--28\,\%) but statistically significant ($p{<}0.01$,
Wilcoxon signed-rank test), confirming generalisation beyond
superconducting hardware. The Adversarial profile retains a 31--45\,\%
advantage, demonstrating robustness. End-to-end latency increases by
1.7--2.1$\times$; all latencies remain below one second, compatible
with quantum cloud pre-processing pipelines.
 
\begin{table}[t]
\centering
\caption{Benchmark summary: SABRE baseline vs.\ joint co-design (Superconducting profile). $S$ and $R$ are means with 95\,\% bootstrap CIs; latency is median end-to-end. All joint gains are significant at $p{<}0.01$ (Wilcoxon signed-rank test, $n{=}100$ paired trials).}
\label{tab:benchmark}
\resizebox{\columnwidth}{!}{%
\begin{tabular}{lcccccc}
\toprule
\multirow{2}{*}{\textbf{Circuit}} &
\multicolumn{3}{c}{\textbf{Baseline (SABRE)}} &
\multicolumn{3}{c}{\textbf{Joint Co-design}} \\
\cmidrule(lr){2-4}\cmidrule(lr){5-7}
& $S$ & $R$ & Lat.\,(ms) & $S$ (95\,\%\,CI) & $R$ (95\,\%\,CI) & Lat.\,(ms) \\
\midrule
VQE-H$_2$  & 0.38 & 0.88 & 180 & 0.64\,(0.60,0.67) & 0.64\,(0.60,0.68) & 370 \\
VQE-LiH    & 0.29 & 0.71 & 413 & 0.55\,(0.51,0.59) & 0.48\,(0.44,0.52) & 720 \\
QPE-12     & 0.22 & 0.62 & 891 & 0.48\,(0.44,0.52) & 0.35\,(0.31,0.39) & 1482\\
Grover-10  & 0.34 & 0.78 & 534 & 0.58\,(0.54,0.62) & 0.52\,(0.48,0.56) & 913 \\
\bottomrule
\end{tabular}}
\end{table}
 
\subsection{Latency vs.\ Success Improvement}
 
\begin{figure}[t]
\centering
\includegraphics[width=0.80\columnwidth]{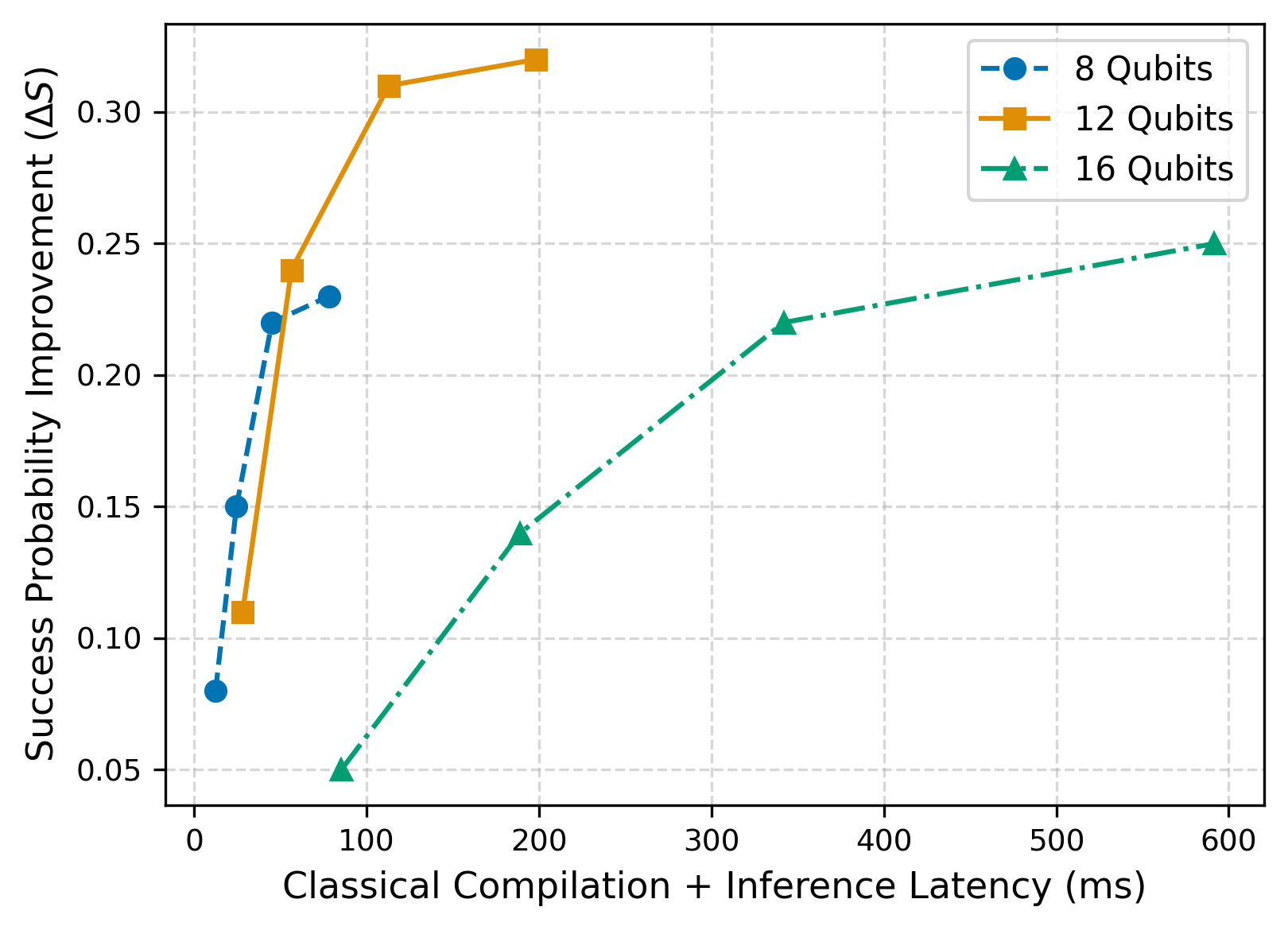}
\caption{Success-probability improvement $\Delta S$ as a function of total compilation and inference latency for 8, 12, and 16 qubits. The diminishing-returns knee near 100\,ms defines a practical upper compilation budget for smaller circuits.}
\label{fig:latency}
\end{figure}

Fig.~\ref{fig:latency} reveals diminishing returns as compilation
budget grows: the marginal success gain falls sharply once latency
exceeds roughly 100\,ms for 8--12 qubit circuits and levels off above
roughly 350\,ms for 16-qubit instances.
 
\subsection{Retention-Success Pareto Analysis}
 
\begin{figure}[t]
\centering
\includegraphics[width=0.80\columnwidth]{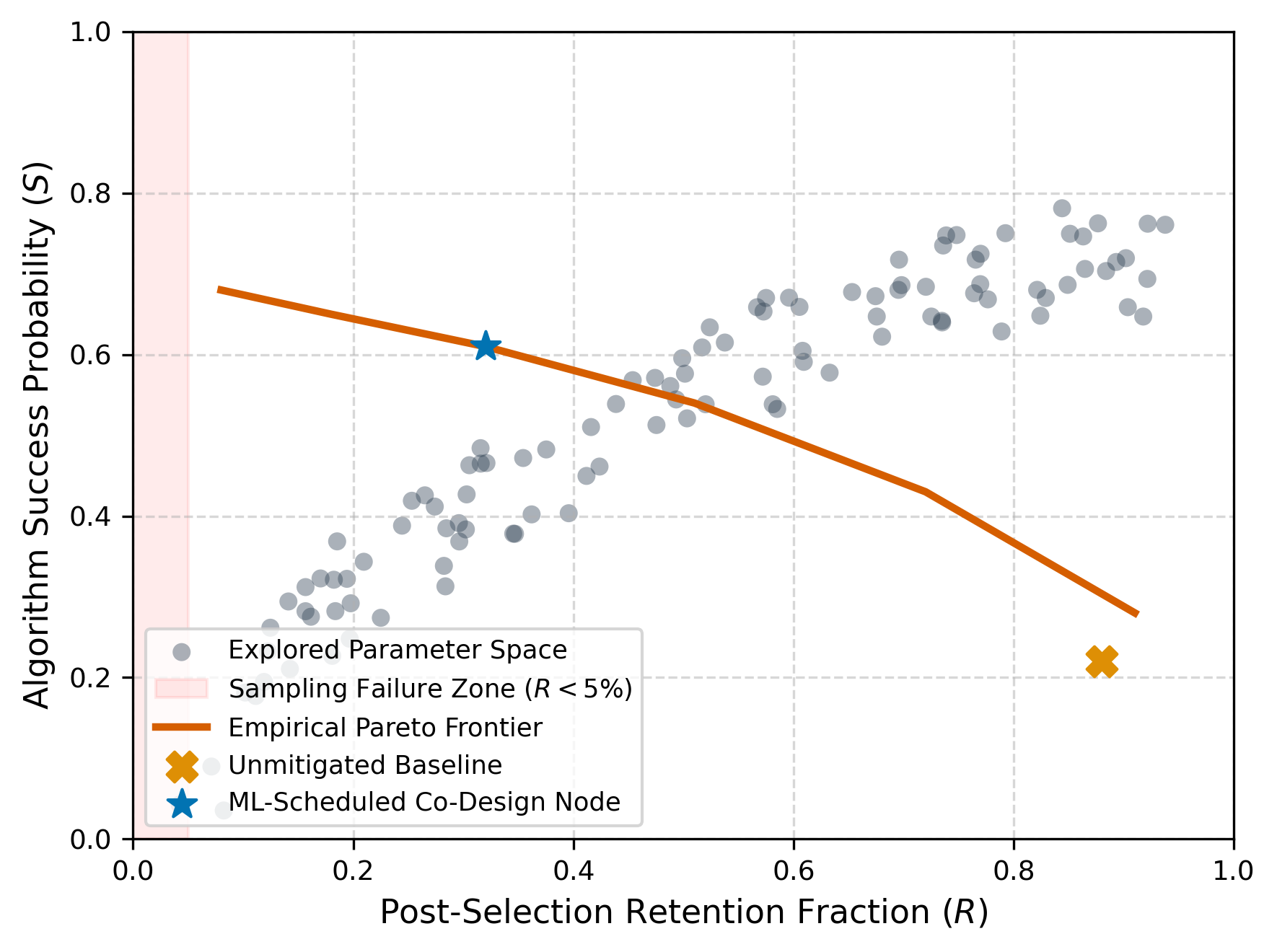}
\caption{Retention-success Pareto scatter. The empirical Pareto frontier (orange) separates feasible from infeasible configurations. The ML co-design point (star) lies on the frontier; the unmitigated baseline (cross) is strictly sub-optimal.}
\label{fig:pareto}
\end{figure}

The joint retention-success Pareto frontier, plotted in
Fig.~\ref{fig:pareto}, reveals the shape of the achievable trade-off
space. The unmitigated SABRE baseline occupies a region of high
retention but low success probability, lying below the Pareto
frontier. The ML-scheduled co-design configuration ($R{=}0.32$,
$S{=}0.61$) falls directly on the empirical frontier. Any
configuration with $R{<}5\%$ is impractical regardless of per-shot
success probability.
 
\subsection{Ablation Study}
 
\begin{figure}[t]
\centering
\includegraphics[width=0.80\columnwidth]{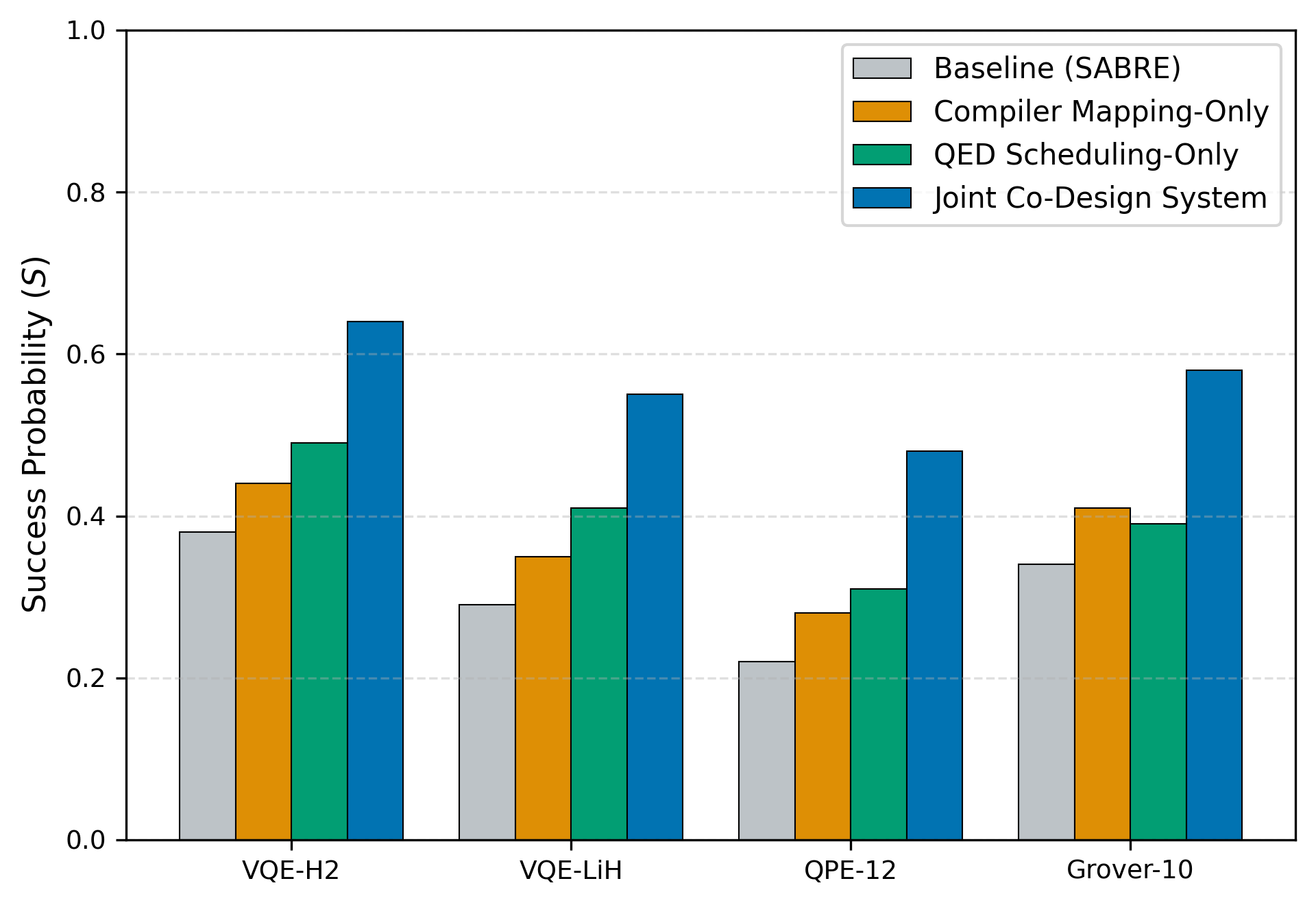}
\caption{Per-benchmark ablation. Joint co-design surpasses both mapping-only and QED-only on every circuit.}
\label{fig:ablation}
\end{figure}

Per-benchmark ablation results are shown in Fig.~\ref{fig:ablation}
and tabulated in Table~\ref{tab:ablation}. For VQE-H$_2$, the mapping
pass alone contributes $\Delta S{=}{+}0.06$ and the QED scheduler
alone $\Delta S{=}{+}0.11$; their joint application yields
$\Delta S{=}{+}0.26$, exceeding the sum (a super-additive gain of
$+0.09$). This arises because noise-aware routing reduces the residual
error presented to the QED layer, improving syndrome reliability. For
Grover-10, the QED-only configuration under-performs mapper-only;
the homogeneous connectivity of this 10-qubit circuit limits targeted
syndrome benefit.
 
\begin{table}[t]
\centering
\caption{Mean success probability $S$ under all ablation configurations (Superconducting profile). Entries are averages over at least 100 trials.}
\label{tab:ablation}
\resizebox{\columnwidth}{!}{%
\begin{tabular}{lcccc}
\toprule
\textbf{Benchmark} & \textbf{Baseline} & \textbf{Mapper-only}
  & \textbf{QED-only} & \textbf{Joint} \\
\midrule
VQE-H$_2$  & 0.38 & 0.44 & 0.49 & \textbf{0.64} \\
VQE-LiH    & 0.29 & 0.35 & 0.41 & \textbf{0.55} \\
QPE-12     & 0.22 & 0.28 & 0.31 & \textbf{0.48} \\
Grover-10  & 0.34 & 0.41 & 0.39 & \textbf{0.58} \\
\bottomrule
\end{tabular}}
\end{table}
 
\subsection{Scaling and Depth Sensitivity}
 
\begin{figure}[t]
\centering
\begin{subfigure}[b]{0.49\columnwidth}
  \centering
  \includegraphics[width=\textwidth]{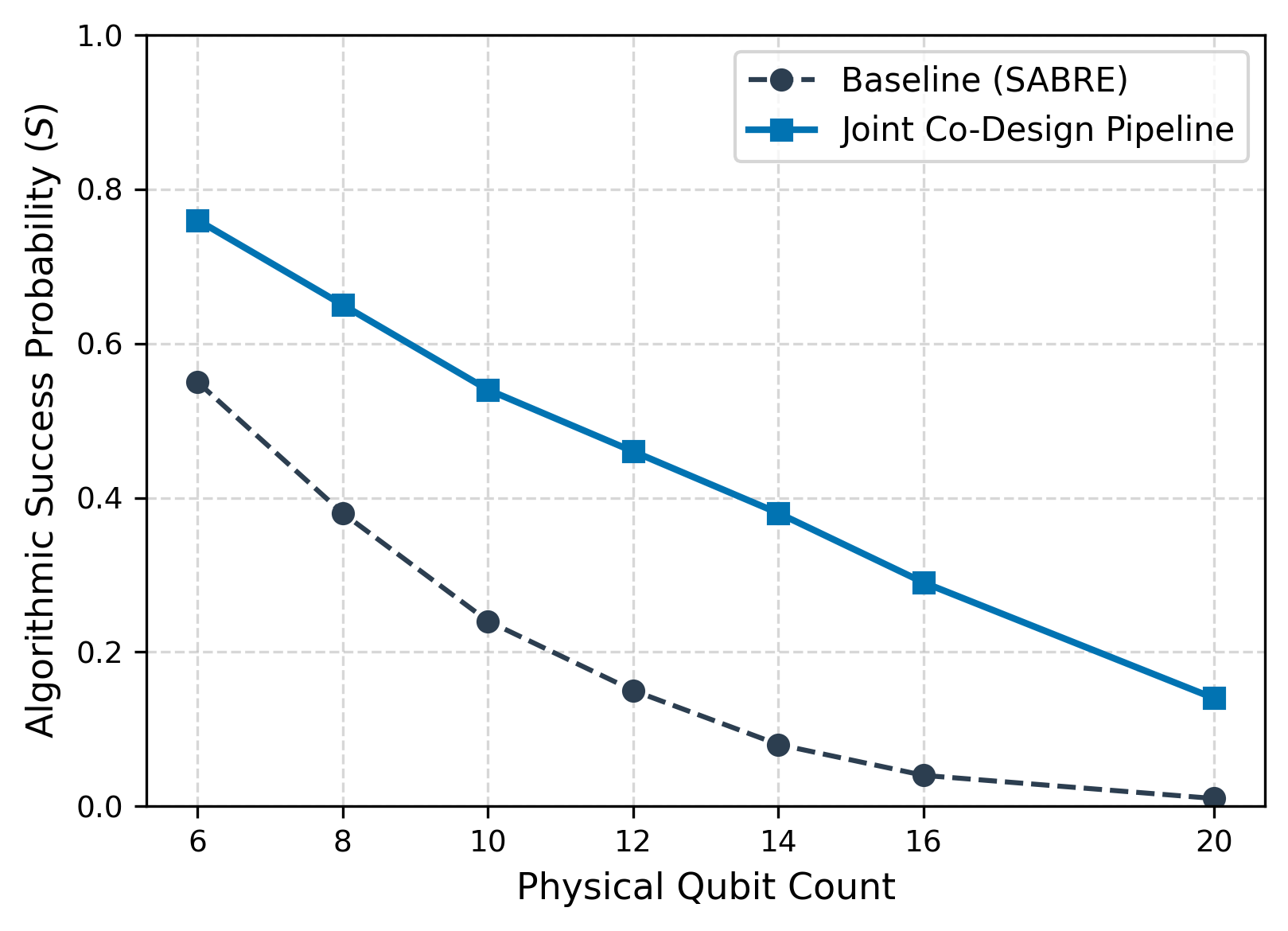}
  \caption{Success vs.\ qubit count.}
\end{subfigure}
\hfill
\begin{subfigure}[b]{0.49\columnwidth}
  \centering
  \includegraphics[width=\textwidth]{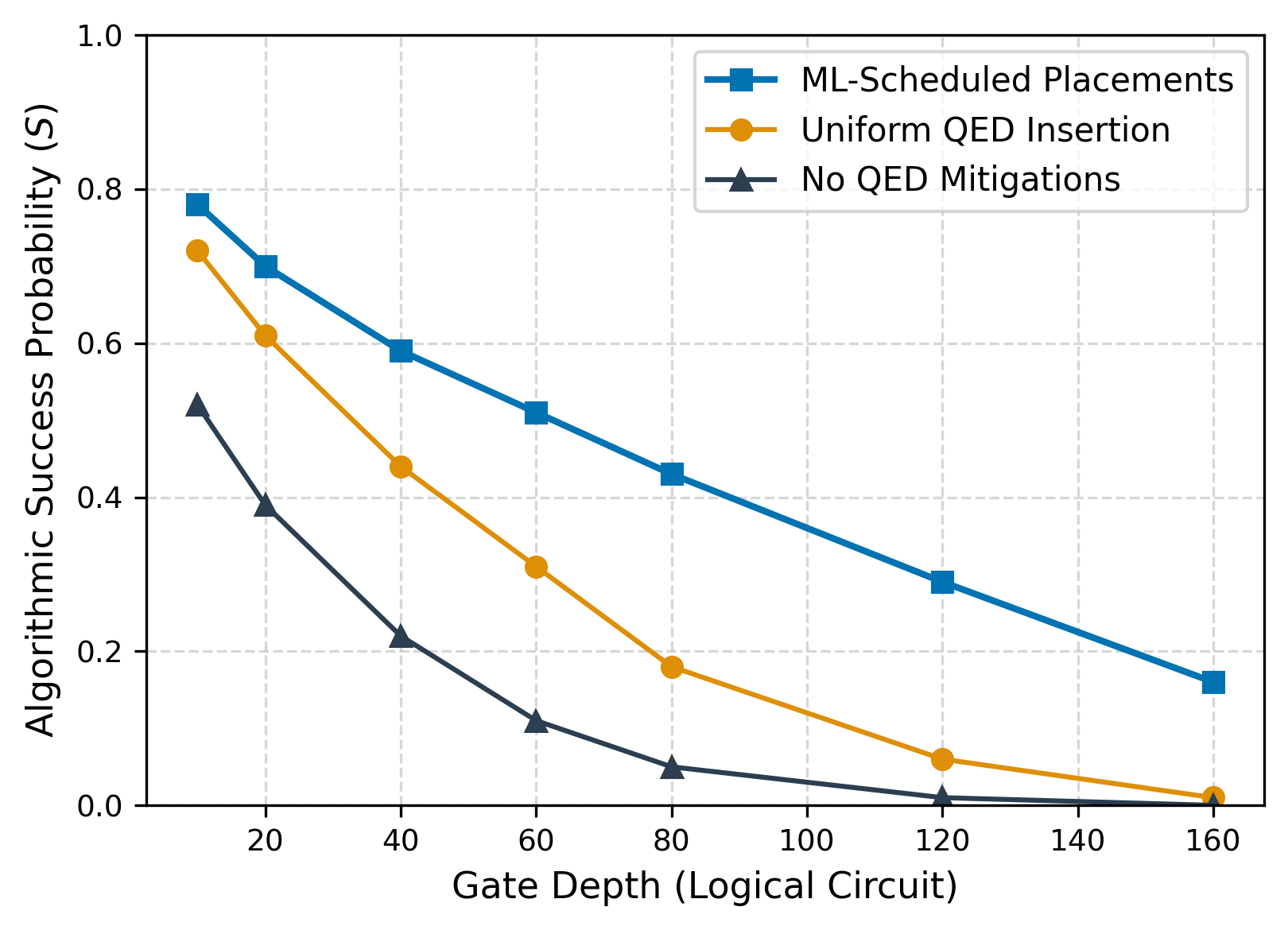}
  \caption{Success vs.\ gate depth.}
\end{subfigure}
\caption{Scaling behaviour. \textbf{(a)}~Success vs.\ qubit count (6--20). \textbf{(b)}~Success vs.\ gate depth (10--160) for ML-scheduled, uniform QED, and no-QED.}
\label{fig:scaling_depth}
\end{figure}

Qubit-count scaling is shown in the left panel of
Fig.~\ref{fig:scaling_depth}: the joint pipeline holds a $>2\times$
advantage over SABRE across the full 6--20 qubit range.
Depth sensitivity appears in the right panel: at gate depth 160, the
unscheduled baseline almost completely fails ($S{\approx}0.01$),
whereas ML-guided syndrome placement sustains $S{\approx}0.16$.
 
\subsection{Overhead, Retention, and Runtime}
 
\begin{figure}[t]
\centering
\begin{subfigure}[b]{0.49\columnwidth}
  \centering
  \includegraphics[width=\textwidth]{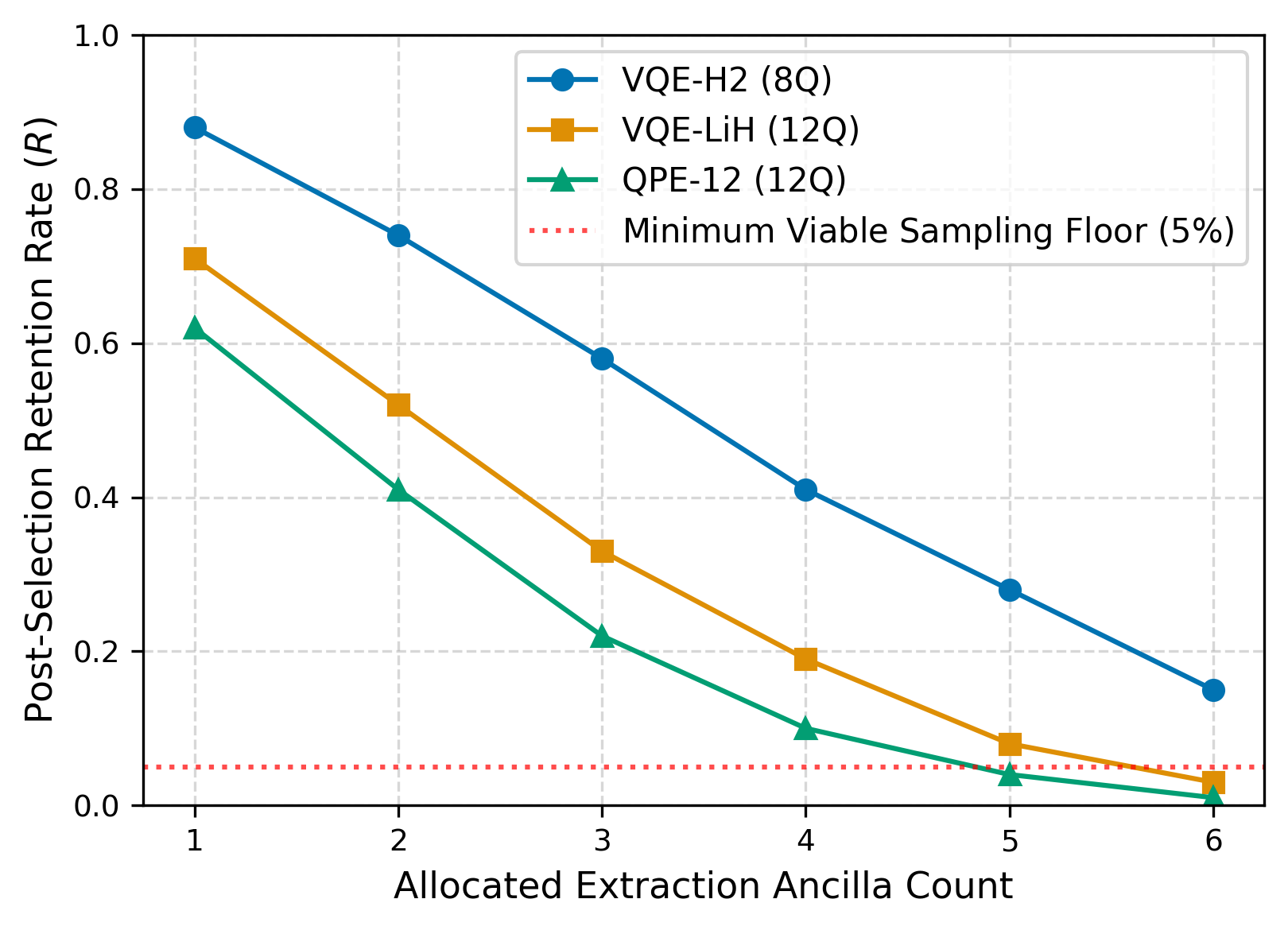}
  \caption{Retention vs.\ ancilla count.}
\end{subfigure}
\hfill
\begin{subfigure}[b]{0.49\columnwidth}
  \centering
  \includegraphics[width=\textwidth]{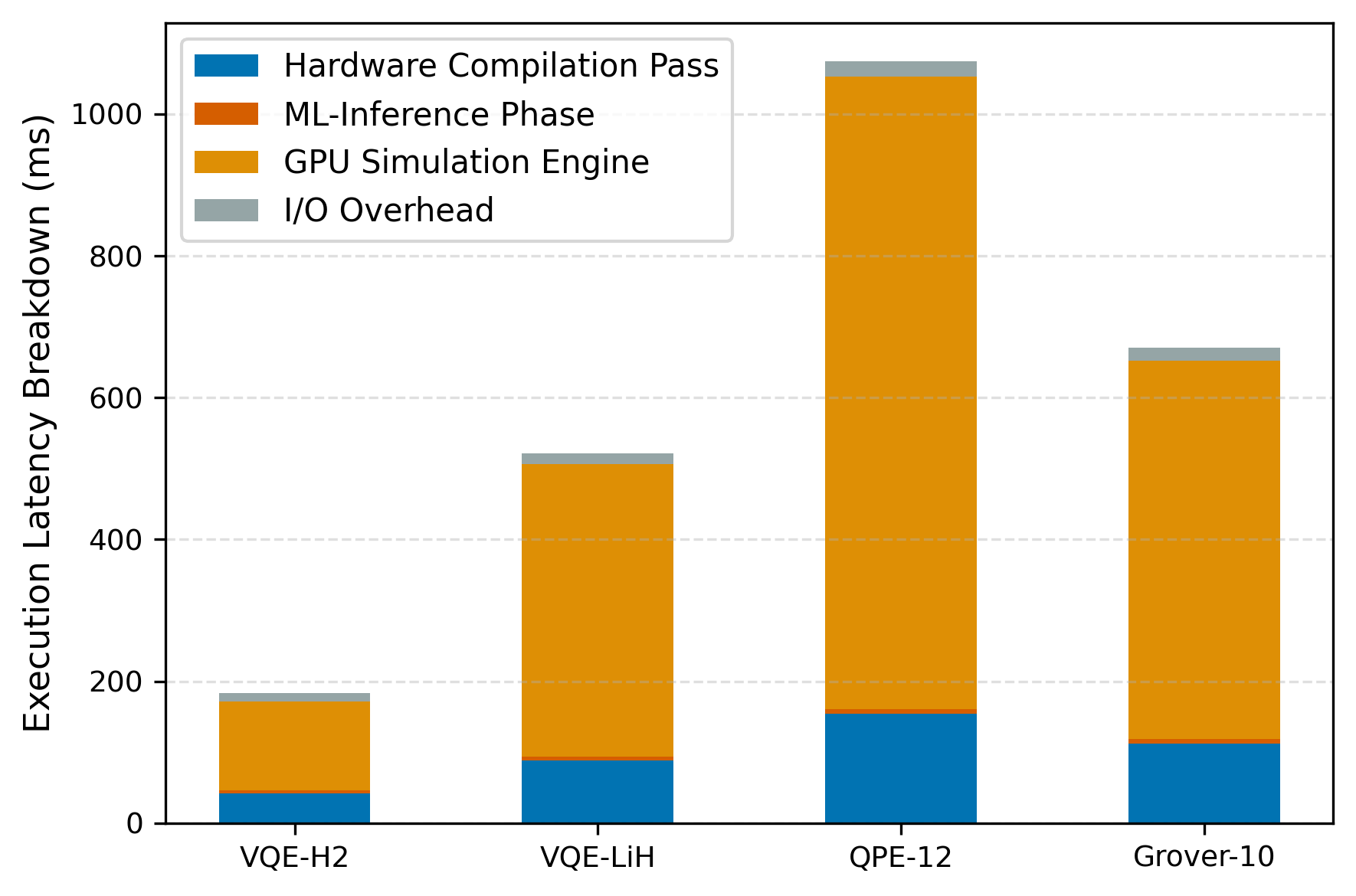}
  \caption{Runtime decomposition.}
\end{subfigure}
\caption{\textbf{(a)}~Retention vs.\ ancilla count; red dashed line: 5\,\% minimum viable floor. \textbf{(b)}~Latency breakdown; GPU simulation dominates.}
\label{fig:overhead_runtime}
\end{figure}

Post-selection retention as a function of ancilla count is shown in
the left panel of Fig.~\ref{fig:overhead_runtime}. For QPE-12,
retention crosses the 5\,\% minimum viable sampling floor at four
ancillae, confirming that the ancilla budget constraint is a practical
necessity. The right panel decomposes total latency: GPU simulation
accounts for 68--83\,\% of wall-clock time; ML inference contributes
at most 1\,\%, confirming negligible classical co-design overhead.
 
\subsection{ML Scheduler Diagnostics}
 
\begin{figure}[t]
\centering
\begin{subfigure}[b]{0.49\columnwidth}
  \centering
  \includegraphics[width=\textwidth]{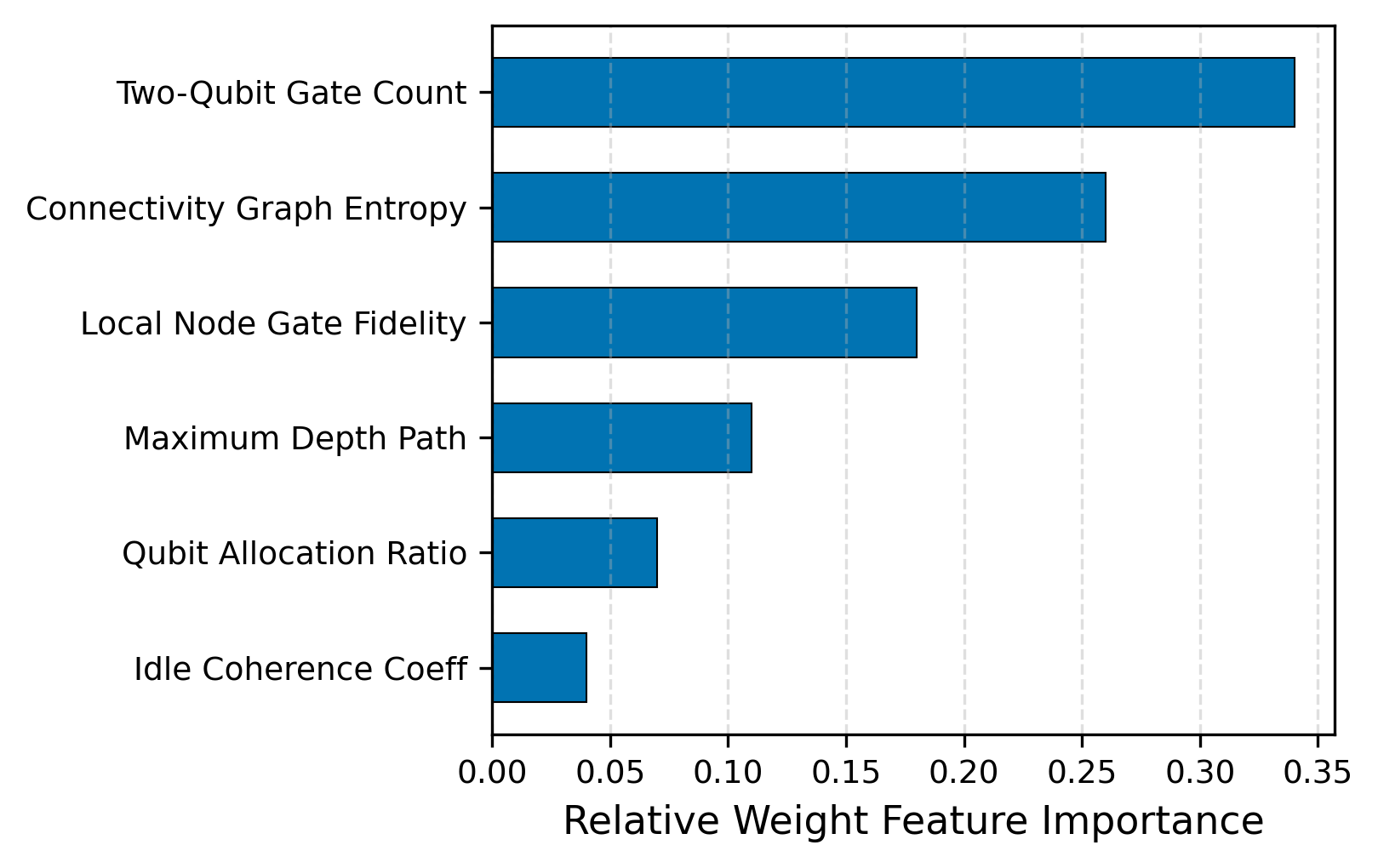}
  \caption{Feature importances.}
\end{subfigure}
\hfill
\begin{subfigure}[b]{0.49\columnwidth}
  \centering
  \includegraphics[width=\textwidth]{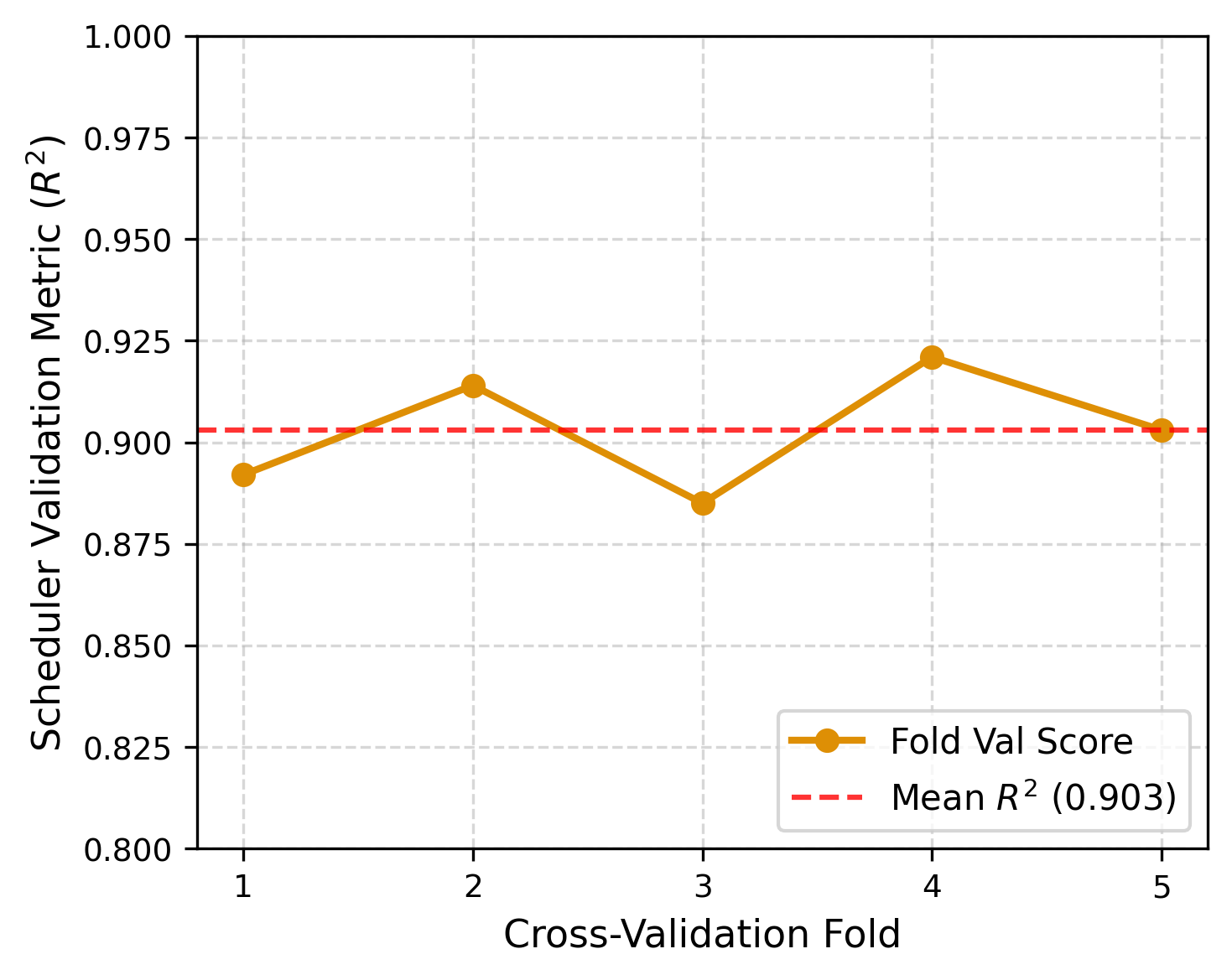}
  \caption{Cross-validation $R^2$.}
\end{subfigure}
\caption{ML diagnostics. \textbf{(a)}~Feature importances for $M_{\Delta S}$. \textbf{(b)}~5-fold CV $R^2$; mean $R^2{=}0.903$.}
\label{fig:ml_diagnostics}
\end{figure}

Two-qubit gate count (importance 0.34) and connectivity entropy (0.26)
are the dominant predictors for $M_{\Delta S}$, accounting for 60\,\%
of predictive variance (Fig.~\ref{fig:ml_diagnostics}a). Five-fold
cross-validation yields mean $R^2{=}0.903$ (range 0.885--0.921),
confirming reliable generalisation (Fig.~\ref{fig:ml_diagnostics}b).
 
\subsection{ILP Kernel Sensitivity}
\label{sec:ilp_sensitivity}
 
Table~\ref{tab:ilp} reports ILP kernel sensitivity.
Performance saturates at $w{=}10$; $w{=}15$ yields no significant
improvement ($p{=}0.71$, Wilcoxon) while adding 65\,\% latency.
The setting $w{=}5$ shows a significant drop ($p{=}0.03$), confirming
$w{\ge}10$ is necessary.
 
\begin{table}[t]
\centering
\caption{ILP kernel sensitivity: mean joint success probability $S$
and end-to-end latency for VQE-H$_2$ (Superconducting profile).}
\label{tab:ilp}
\resizebox{0.85\columnwidth}{!}{%
\begin{tabular}{cccc}
\toprule
\textbf{$w$} & \textbf{$S$ (Joint)} & \textbf{95\,\% CI} & \textbf{Lat.\,(ms)} \\
\midrule
5  & 0.61 & (0.57, 0.65) & 290 \\
\textbf{10} & \textbf{0.64} & \textbf{(0.60, 0.67)} & \textbf{370} \\
15 & 0.64 & (0.60, 0.68) & 610 \\
\bottomrule
\end{tabular}}
\end{table}
 
\subsection{Simulator vs.\ Hardware Validation}
\label{sec:hwval}
 
\begin{figure}[t]
\centering
\includegraphics[width=0.80\columnwidth]{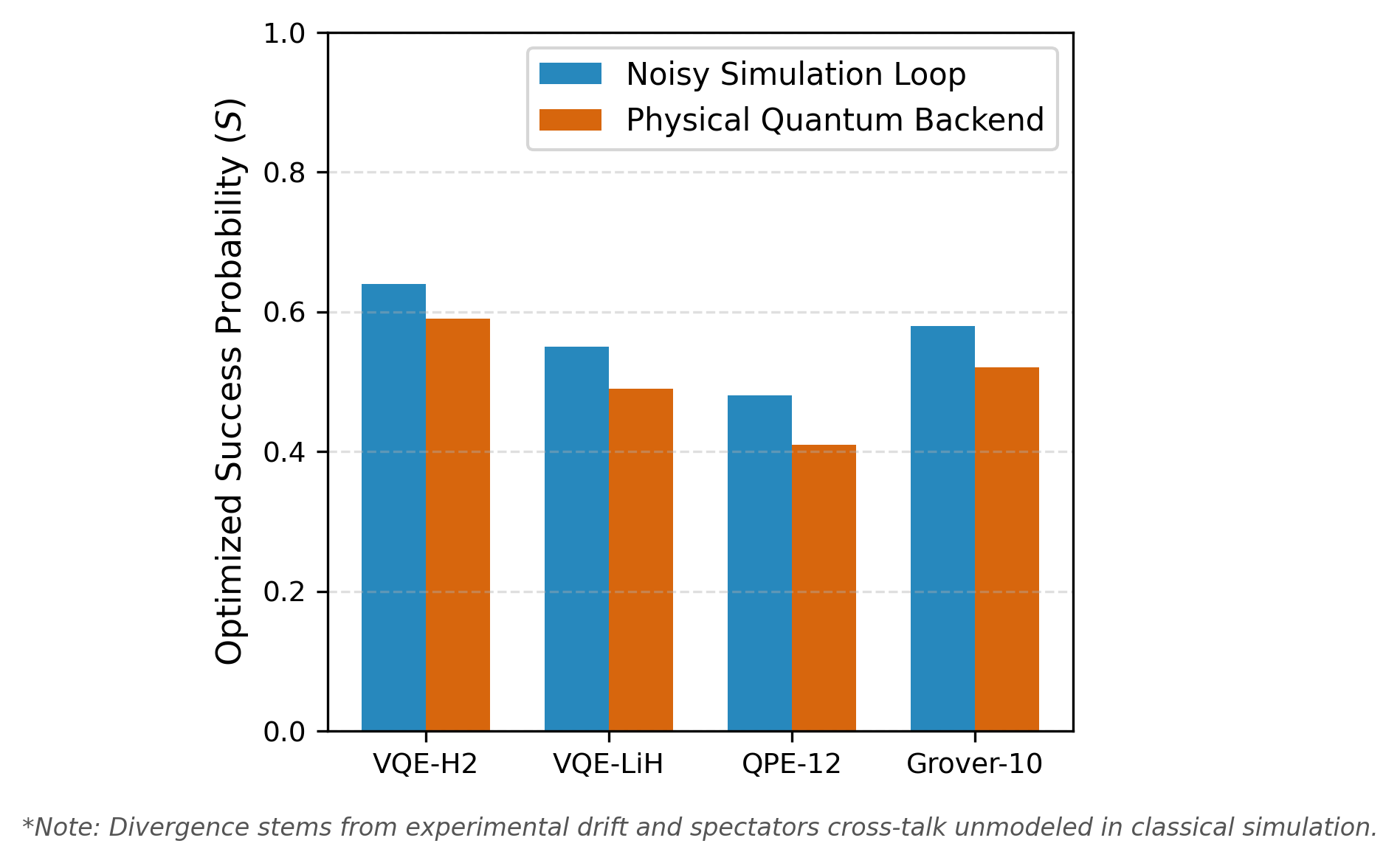}
\caption{Simulation vs.\ IBM Eagle hardware (8,192 shots). The 4--8\,pp gap reflects unmodelled drift and cross-talk; directional rankings are preserved across both hardware-feasible benchmarks.}
\label{fig:hw}
\end{figure}

Fig.~\ref{fig:hw} compares optimised success probabilities from
classical simulation against results from 8\,192-shot runs on an IBM
Eagle-family superconducting processor (Table~\ref{tab:hwmeta}).
Hardware results fall consistently 4--8 percentage points below
simulation across both VQE-H$_2$ and Grover-10, attributable to
unmodelled drift, spectator cross-talk, and calibration staleness.
The \emph{relative rankings} of configurations and the
\emph{directional advantage} of joint co-design over SABRE are
preserved in both settings. Per-configuration hardware ablation across
all benchmarks remains outside present IBM Quantum queue constraints.
 
\section{Discussion}
\label{sec:discussion}
 
The ablation results establish a super-additive interaction between
the two co-design components: routing quality constrains the residual
error that QED must absorb, so a better mapping improves syndrome
reliability and amplifies the net utility of error detection. Classical
co-design overhead is negligible relative to quantum simulation time.
Proposition~\ref{prop:cost} grounds the cost function in first-order
Depolarising infidelity; the SA+ILP search has no worst-case polynomial
guarantee, and $\alpha$ and $w$ jointly govern the quality-time
trade-off, confirmed empirically across all benchmarks
(Table~\ref{tab:ilp}).
 
\textbf{Simulation-to-hardware gap.}
The 4--8\,pp gap (Fig.~\ref{fig:hw}) has three contributors:
(a)~\emph{calibration drift} (1--3\,pp) between snapshot and execution;
(b)~\emph{spectator ZZ cross-talk} (1--4\,pp) absent from the
independent-Depolarising model~\cite{Nation2023};
(c)~\emph{heterogeneous SPAM errors} varying up to $2\times$ per qubit.
Drift-aware noise modelling and SPAM correction are the highest-priority improvements.
 
\textbf{Limitations.}
\textit{(i)}~Primary results are simulation-based; absolute values require hardware re-calibration.
\textit{(ii)}~The ML scheduler requires retraining after significant device drift.
\textit{(iii)}~Post-selection retention drops sharply with ancilla count; shot budgets must be sized accordingly.
\textit{(iv)}~Exact simulation is bounded to 6--20 qubits; tensor-network approximations add error at larger scales.
\textit{(v)}~Hardware validation covers only VQE-H$_2$ and Grover-10; full multi-benchmark ablation is pending.
\textit{(vi)}~The XGBoost scheduler does not incorporate topological graph structure; a graph neural network remains an open extension.
 
\section{Conclusion}
\label{sec:conclusion}
 
We have presented an integrated hardware-aware compilation and
data-driven QED framework for the early fault-tolerance regime.
Unifying qubit mapping, SWAP insertion, and syndrome-schedule placement
under a single noise-weighted objective yields up to 68\,\% improvement
in algorithmic success probability (95\,\% CI: 60\,\%--76\,\%) over
SABRE across all four benchmarks and three noise profiles. On VQE-H$_2$,
the super-additive gain ($\Delta S_\mathrm{joint}{=}0.26 >
\Delta S_\mathrm{mapper}{+}\Delta S_\mathrm{QED}{=}0.17$) demonstrates
co-design synergy inaccessible to either component alone. The cost
function is grounded as a first-order Depolarising infidelity bound
(Proposition~\ref{prop:cost}), the noise-adaptive literature is subsumed
as a special case ($w_g{=}1$), and the full pipeline is reproducible via
the released Docker and SLURM infrastructure. Future work targets
drift-aware noise modelling, SPAM correction, and higher-distance
stabiliser codes.
 
\begin{credits}
\subsubsection{\ackname}
The author thanks the Quantum Technology group at IIT Jodhpur for
discussions; the IIT Jodhpur HPC Centre for computational resources;
IBM Quantum for Eagle-family processor access; and NVIDIA for the cuQuantum SDK.
\subsubsection{\discintname}
The author declares no competing interests.
\end{credits}
 
\paragraph{Data Availability.}
Source code and SLURM/Docker scripts are available at
\url{https://github.com/Sumitchongder/quantum-hw-aware-pipeline}.
 

\end{document}